\documentstyle[11pt,paspconf,epsf]{article}

\begin{document}

\title{Aperture Synthesis Images of Dense Molecular Gas
    in Nearby Galaxies with the Nobeyama Millimeter Array}

\author{K. Kohno, R. Kawabe, T. Shibatsuka, and S. Matsushita}
\affil{Nobeyama Radio Observatory, Minamisaku, Nagano, 384-1305, Japan}




\begin{abstract}
High resolution images of HCN(1$-$0) and HCO$^+$(1$-$0) 
emissions from nearby galaxies made with the NMA are presented.
\end{abstract}


\keywords{galaxies, starburst, Seyfert, HCN, HCO$^+$, dense molecular gas}


\section{Dense Molecular Gas in Galaxies}

In order to study the distribution of dense molecular gas 
and its relation to the central activities (starburst and AGN) in galaxies,
we have conducted an imaging survey of HCN(1$-$0) and HCO$^+$(1$-$0) emissions
from nearby spiral galaxies with the Nobeyama Millimeter Array (NMA)
(Kohno et al.\ 1996, 1998, 1999a, 1999b, 1999c; Shibatsuka et al.\ 1999).
Figure 1 shows preliminary images of HCN and HCO$^+$ in galaxies.
In starburst galaxies, we find there is good
spatial coincidence between dense molecular gas and star-forming
regions. The ratios of HCN to CO integrated intensities on the brightness
temperature scale, $R_{\rm HCN/CO}$, are as high as 0.1 to 0.2 in the
starburst regions, and quickly decrease outside of these regions. 
In contrast, we find a remarkable decrease of 
the HCN emission in the post-starburst nuclei, despite the strong
CO concentrations there. The $R_{\rm HCN/CO}$ values in the
central a few 100 pc regions of these quiescent galaxies are very low,
0.02 to 0.04. A rough correlation between $R_{\rm HCN/CO}$ and
H$\alpha$/CO ratios, which is an indicator of star-formation efficiency,
is found at a few 100 pc scale.
The fraction of {\it dense} molecular gas
in the {\it total} molecular gas, measured from $R_{\rm HCN/CO}$,
may be an important parameter that controls star formation.
In some Seyfert galaxies we find extremely high
$R_{\rm HCN/CO}$ exceeding 0.3. These very high ratios are
never observed even in strong starburst regions,
implying a physical link between extremely high $R_{\rm HCN/CO}$
and Seyfert activity.

%
%
%
 
%


\begin{figure}
\plotone{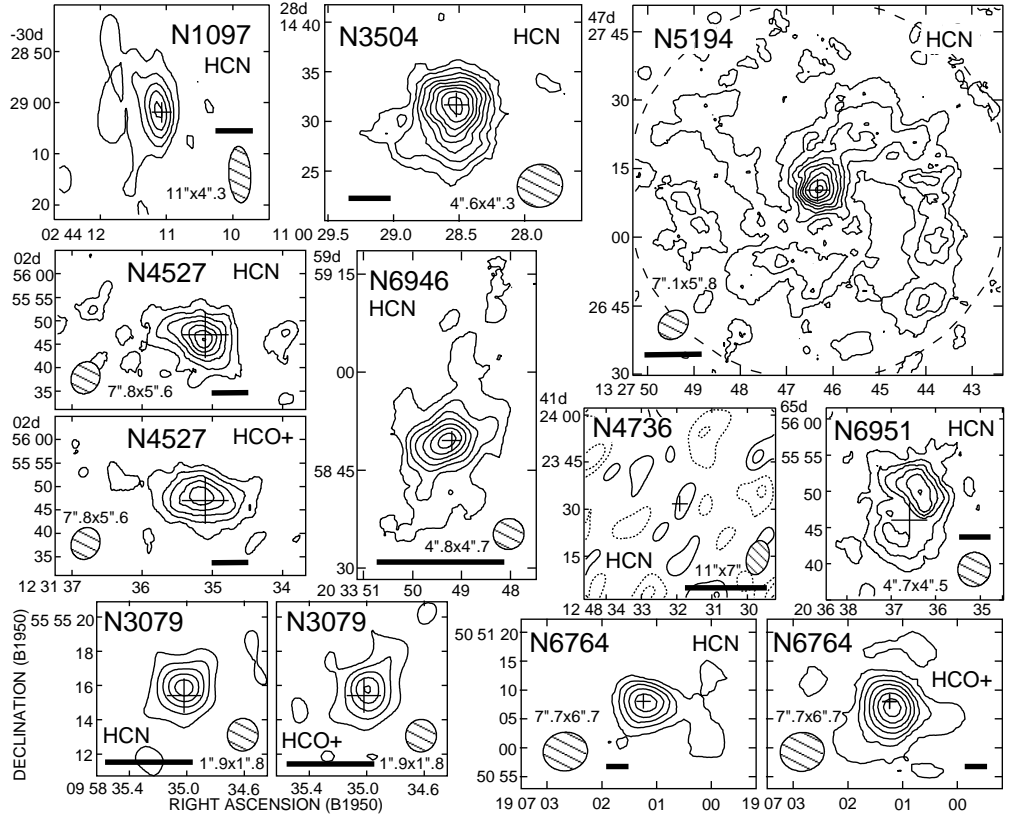}
\caption{High resolution HCN and HCO$^+$ images of nearby 
spiral galaxies. HCO$^+$ images were obtained simultaneously
with the Ultra Wide-Band Correlator of the NMA (1024 MHz
band-width).
The thick horizontal line in each image corresponds to 500 pc.
The cross marks the position of a 6 cm radio continuum peak.
The contour interval is 2 $\sigma$
for NGC 3079, NGC 3504, NGC 6764, and NGC 6946,
and 1.5 $\sigma$ for NGC 1097, NGC 4527, NGC 5194, and NGC 6951.
Contour levels are -3, -1.5, 1.5, and 3 $\sigma$ for NGC 4736.
Noise levels are 1.7, 1.5, 0.46, 0.71, 0.91, 0.44, 0.36, 0.83, 
and 0.58 Jy beam$^{-1}$ km s$^{-1}$
for NGC 1097, NGC 3079, NGC 3504, NGC 4527, NGC 4736, 
NGC 5194, NGC 6764, NGC 6946, and NGC 6951, respectively.
A {\it nuclear} starburst is hosted by 
NGC 3079, NGC 3504, NGC 4527, NGC 6764, and NGC 6946,
whereas a circumnuclear star-forming 
{\it ring} is observed in NGC 6951. Good spatial coincidence between
HCN and H$\alpha$ (and radio continuum) emission can be seen 
in these star-forming galaxies (Kohno et al. 1999a).
Note that there is a remarkable decrease of dense gas 
in the post-starburst galaxy NGC 4736; 
the current massive star formation is suppressed 
in the central $r \sim 10''$ region, 
even though there is a CO peak. The resultant
$R_{\rm HCN/CO}$ is very small, less than 0.044 (2 $\sigma$ upper limit).
The sample contains one type-1 Seyfert (NGC 1097) 
and three type-2 Seyferts (NGC 3079, NGC 5194, and NGC 6951).
A cohabitation of AGN and starburst has been reported to have taken place
in NGC 1097, NGC 3079, and NGC 6951.
} \label{fig-1}
\end{figure}

\end{document}